\documentclass[10pt,journal,compsoc]{IEEEtran}

\usepackage{amsmath}
\usepackage{cite}
\usepackage{graphicx}
\usepackage{footnote}
\usepackage{bbding}
\usepackage{pifont}
\usepackage{wasysym}
\usepackage{amssymb}
\usepackage{multirow}
\usepackage[table,xcdraw]{xcolor}
\usepackage{diagbox}
\usepackage[flushleft]{threeparttable}
\makesavenoteenv{tabular}
\usepackage{tabu}
\usepackage{longtable}
\usepackage{cite}
\usepackage{epstopdf}
\usepackage{color}
\usepackage{enumitem}
\usepackage{booktabs}
\usepackage[noend]{algpseudocode}
\usepackage[ruled]{algorithm2e}
\usepackage{subfig}
\usepackage{longtable}

\setlist{parsep=0pt,listparindent=\parindent}

\ifCLASSINFOpdf

\else

\fi

\usepackage{ragged2e}
\begin{document}

\title{Link Prediction Adversarial Attack}

\author{Jinyin~Chen, Ziqiang~Shi, Yangyang~Wu, Xuanheng~Xu and Haibin~Zheng
\IEEEcompsocitemizethanks{
\IEEEcompsocthanksitem J. Chen, J. Shi, Y. Wu, X. Xu and H. Zheng are with the College of Information Engineering, Zhejiang University of Technology, Hangzhou 310023, China\protect\\

E-mail: \{chenjinyin, 201603090114, 2111603080, 2111603112, 201303080231\}@zjut.edu.cn
\IEEEcompsocthanksitem Corresponding author: Jinyin~Chen
}
}


\IEEEtitleabstractindextext{
\begin{abstract}
\justifying
Deep neural network has shown remarkable performance in solving computer vision and some graph evolved tasks, such as node classification and link prediction. However, the vulnerability of deep model has also been revealed by carefully designed adversarial examples generated by various adversarial attack methods. With the wider application of deep model in complex network analysis, in this paper we define and formulate the link prediction adversarial attack problem and put forward a novel iterative gradient attack (IGA) based on the gradient information in trained graph auto-encoder (GAE). To our best knowledge, it is the first time link prediction adversarial attack problem is defined and attack method is brought up. Not surprisingly, GAE was easily fooled by adversarial network with only a few links perturbed on the clean network. By conducting comprehensive experiments on different real-world data sets, we can conclude that most deep model based and other state-of-art link prediction algorithms cannot escape the adversarial attack just like GAE. We can benefit the attack as an efficient privacy protection tool from link prediction unknown violation, on the other hand, link prediction attack can be a robustness evaluation metric for current link prediction algorithm in attack defensibility.

\justifying
\end{abstract}

}

\maketitle

\IEEEdisplaynontitleabstractindextext

\ifCLASSOPTIONcompsoc
\IEEEraisesectionheading{\section{Introduction}\label{sec:introduction}}
\else
\section{Introduction}
\label{sec:introduction}
\fi

\IEEEPARstart{M}{a}{n}{y}
real-world systems can be represented as a network, such as social networks~\cite{Borgatti2009Network, Wellman2005The}, biological networks~\cite{Montoya2002Small}, communication networks~\cite{Ebel2002Scale}, traffic networks~\cite{Latora2002Is} and so on. In particular, some systems especially the social networks are dynamic, and the links in such network always change over time~\cite{Neal2008, Liben2003The}. How to test out the undiscovered link or the link which will change in the future are usually known as the link prediction problem. Link prediction is capable to benefit a wide range of real-world applications. For instance, if we have already known some terrorist communications, we may find out some hidden links so as to discover the potential terrorists by link prediction~\cite{Anil2016Link}. Link prediction can also serve to recommendation systems~\cite{chen2017improved,chen2017double, Huang2005Link, Yang2012Scalable}, network reconstruction~\cite{Guimer2009Missing}, and node classification~\cite{Bilgic2007Combining}, etc.

In the past few decades, lots work of link predictions have been proposed. The similarity-based algorithm is one of the most popular link prediction methods, which assumes that the more similar two nodes are, the more likely they are to be linked. Node similarity can be directly captured by the essential attributes of nodes, such as the personal information in social network~\cite{Lin1998An}, paper detail in citation network~\cite{Popescul2003Statistical} and semantic information of web page network~\cite{Craven1998Learning}. However, since the content and attribute information are generally not available due to limitations, most cases focus on the structural similarity only.
Neighbor-based similarity method is a simple but efficient similarity-based method. The Common Neighbors (CN) and Resource Allocation (RA)~\cite{Zhou2009Predicting}, classic neighbor-based similarity methods, successfully applied to personalized recommendation~\cite{Zhou2007Bipartite},~\cite{L2011Link}.
Path-based index and random-walk-based index~\cite{Liu2010Link} are two other kinds of similarity index, taking advantage of network's global topological information, usually perform better, such as the Katz index~\cite{Katz1953A} and Local Random Walk (LRW)~\cite{Liu2010Link}.
Recently, with the
tremendous development of deep learning models, the embedding models have achieved remarkable performance in many network tasks. Inspired by the advancements of language model word2vec~\cite{Mikolov2013Efficient}, some unsupervised network embedding methods such as DeepWalk~\cite{Perozzi2014DeepWalk}, LINE~\cite{Tang2015LINE} and node2vec~\cite{Grover2016node2vec} have achieved great success. The embedding learned by these models can be directly applied to a variety of network tasks including link prediction~\cite{Perozzi2014DeepWalk}, graph classification~\cite{Bunke2008Graph, Riesen2010Graph}, node classification~\cite{Tang2015PTE, Wang2016Linked}, and community detection~\cite{Tian2014Learning, Allab2016A}.
The graph auto-encoder model for link prediction, proposed by Kipf et al.~\cite{Kipf2016Variational}, is a variation of GCN which can learn  node representation efficiently by layer-wise propagation rule.
Due to their non-linear and hierarchical nature, deep learning methods were shown great power and large potential in link prediction~\cite{Grover2016node2vec ,Schlichtkrull2017Modeling}. However, on the other hand, the vulnerability of deep model was also revealed. In the field of computer vision, deep models are easily fooled by carefully crafted adversarial examples which has slight perturbation on the clean image to make the deep model to achieve an incorrect result.~\cite{Goodfellow2014Explaining, Szegedy2013Intriguing}. The perturbations are usually hard to pick out by human but let the deep model wrongly predict with high confidence~\cite{Goodfellow2014Explaining, Moosavidezfooli2016DeepFool}. The attack triggered by adversarial examples are named adversarial attack. Here we are interested in the questions: can link prediction deep model be attacked by adversarial examples? The non-deep predictors can be attacked as well? We propose and formulate the link prediction adversarial attack problem, put forward a gradient based attack method and discuss the transfer attack on other predictors. The study of this paper can benefit protecting the user privacy for certain cases, and on the other hand, the attack is an efficient tool of robustness evaluation for these state-of-art link prediction methods.

There are interesting works focusing on network safety problem. By adding links to the nodes of high centrality, Nagaraja~\cite{nagaraja2010impact} proposed the community deception method against community detection algorithms. Waniek et al.~\cite{waniek2018hiding} proposed Disconnect Internally, Connect Externally (DICE) algorithms, which can conceal community by adding and removing some specific links. Besides, in the field of node classification, NETTACK was proposed by Zugner et al.~\cite{DBLP:conf/kdd/ZugnerAG18} to make node classification algorithm invalid. NETTACK generates adversarial network iterative according to the changing of confidence value  after adding perturbations. They further proposed a principled strategy~\cite{bojcheski2018adversarial} for adversarial attacks on unsupervised node embeddings, and demonstrated that the attacks generated by node embedding have significant negative effect on node classification and link prediction. Dai et al.~\cite{dai2018adversarial} proposed a reinforcement learning based attack method, which learns the generalizable attack strategy and only requiring prediction labels from the target classifier.
In aspect of network embedding, Chen et al.~\cite{chen2018fast} proposed the adversarial attack on a variety of network embedding models via gradient of deep model , named FGA. FGA is capable of conducting network embedding adversarial attack by merely modifying a few links to change the embedding of target node significantly.

For link prediction, Zheleva et al.~\cite{zheleva2008preserving} proposed the link re-identification attack to predict sensitive links from the released data. Link perturbation is a common technique in early research that data publisher can randomly modify links on the original network to protect the sensitive links from being identified. Fard et al.~\cite{fard2012limiting} introduced a subgraph-wise perturbation in directed networks to randomize the destination of a link within subgraphs to protect sensitive links. They further proposed a neighborhood randomization mechanism to probabilistically randomize the destination of a link within a local neighborhood~\cite{fard2015neighborhood}. Distinguishing from the perspective of privacy protection, these methods for privacy protection can also be used evilly. For instance, the terrorists in a communication network may use the same technique to hide their activities. Many recommender system
based on link prediction methods may also be fooled, which can make fraudsters or their products to be the one be recommended to innocent users. Therefore, in the necessity of privacy protection and measuring the robustness of link prediction algorithms, the study of link prediction attack on some certain links is urgently need. Until now, to our best knowledge, it is the first time link prediction adversarial attack is proposed.

In this paper, we propose a novel iterative gradient attack (IGA), which is capable of conducting adversarial attack on link prediction. Specifically, we make the following contributions.

\begin{itemize}
\item We define and formulate the link prediction adversarial attack problem, and propose a novel iterative gradient attack (IGA) against GAE model. With slight link perturbation on the original clean network, the adversarial network can lead the GAE to totally incorrect link prediction results.

\item Since real-world networks are complicated and usually encounter limitations when conduct attacks, we propose two adversarial attack strategies, unlimited attack and single node attack. The unlimited attack assumes the attacker has high authority while single node attack is more concealed but challenging.

\item IGA has great attack transiferablitiy against deep or other link prediction methods, including state-of-art deep models and other classic similarity-based algorithms. Comprehensive experiments are carried out to testify its performance on different real-world data sets.

\end{itemize}

The rest of paper is organized as follows.
In Sec.~\ref{Method}, we introduce our IGA method in details and explain its attack ability against GAE. Attack transferability of IGA is analyzed as well.
In Sec.~\ref{Exp}, we empirically evaluate on multiple tasks, and compare the attack effects by utilizing IGA and other baseline attack methods on several real-world networks.
In Sec.~\ref{Conclusion}, we conclude the paper and highlight future research directions.


\section{Method\label{Method}}
In this section, we will define and formulate link prediction attack problem, and then introduce how adversarial network is generated via IGA to conduct attack against deep model based link prediction method, such as GAE, and other classic link prediction methods.

\begin{table}[!t]
\centering
\caption{The definitions of symbols.}
 \label{Definition}
 \resizebox{\linewidth}{!}{
\begin{tabular}{lr}
\hline \hline
Symbol & Definition \\ \hline
   $G$=$(V,E)$           & original network with nodes $V$ and links $E$\\
   $\hat{G}$=$(V,\hat{E})$       & adversarial network with nodes $V$ and updated links $\hat{E}$\\
   $E_o$, $E_u$ and $E_t$        & the observed link set, the unknown link set and  the target link\\
   $E_\beta$, $E_{\beta+}$/$E_{\beta-}$ & the total perturbation link set, the added/removed link set of $G$\\
   $A$ and $\hat{A}$         & the adjacency matrix of $G$ and adversarial network $\hat{G}$\\
   $W_i$, $s$ and $\sigma$      & the weight of $i$th layer, the sigmoid and Relu activation function in GAE\\
   $L$ and $\tilde{L}$     		& the loss function of GAE and the target loss function for $E_t$\\
   $Y_t$         	& the ground truth label of target link $E_t$\\     	
   $\eta$        	& learning rate of training GAE\\
   $\tilde{A}$   	& the score matrix calculated by GAE for link prediction\\
   $g$ and $\hat{g}$& the gradient matrix and the symmetric gradient matrix\\
   $n$           	& the number of link modified in an iteration\\
   $K$           	& the iteration number for gradient calculation\\
   $k_t$         	& the link degree of target link $E_t$\\
   $\hat{A}^h$ 		& the $h^{th}$ adversarial adjacency matrix\\
   $\hat{g}^h$ 		& the $h^{th}$ link gradient network\\
\hline \hline
\end{tabular}}
\end{table}

\subsection{Problem Definition}



\newtheorem{Link prediction}{\textbf{\textsc{Definition}}}
\newtheorem{Adversarial network}[Link prediction]{\textbf{\textsc{Definition}}}
\newtheorem{Link prediction adversarial attack}[Link prediction]{\textbf{\textsc{Definition}}}
\begin{Link prediction}[Link Prediction]
For a given network represented as $G$=$(V,E)$, $V$ denotes the set of all nodes and $E$ denotes the set of all links. $E$ is divided into two groups $E_{o}$, $observable$ by so far information, and $E_{u}$, $unknown$ by now need to be predicted, where $E_{o}\cap E_{u}$=$\phi$ and $E_{o}\cup E_{u}$=$E$. Link prediction is conducted to predict $E_{u}$ based on the information of $V$ and $E_{o}$.

\end{Link prediction}

\begin{Adversarial network}[Adversarial Network]
Given the original clean network $G$=$(V,E)$, adversarial network $\hat{G}$=$(V,\hat{E})$ is constructed by adding negligible perturbations $E_\beta$ on $G$, to make link prediction methods $f$ fail to predict targeted link $E_{t}\in E_u$, formulate as


\begin{equation}
\hat{E}= {E}+E_{\beta+}-E_{\beta-},
\end{equation}

where $E_{\beta+}\cup E_{\beta-} = E_\beta$, $E_{\beta+}\cap E_{\beta-}=\phi$, $E_{\beta-}\subset E_o$, $E_{\beta+}\subset(\Omega-E)$ and $\Omega = \{(i, j), \{i, j\}\in V, i \ne j\}$.
\end{Adversarial network}

\begin{Link prediction adversarial attack}[Link Prediction Adversarial Attack]
For the given network $G$ and target link $E_t$, generate an adversarial network $\hat{G}$ to replace $G$ as the network input to make Link prediction method $f$ fail to make accurate prediction of link $E_t$. The attack can be formulate as


\begin{equation}
\begin{array}{ll}
\max\limits_{\hat{G}} \quad \mathbb I(f( G, E_t ) \ne f( \hat{G}, E_t ))\\ s.t. \quad |E_\beta| \leqslant m,
\end{array}
\end{equation}
where $\mathbb I(\cdot)\in\{0, 1\}$ is an indicator function, $m$ is the maximum number of modified links, $E_\beta=E_{\beta+}\cup E_{\beta-}$.


\end{Link prediction adversarial attack}

\subsection{Framework}
\begin{figure*}[!t]
  \centering
\includegraphics[width=0.9\linewidth]{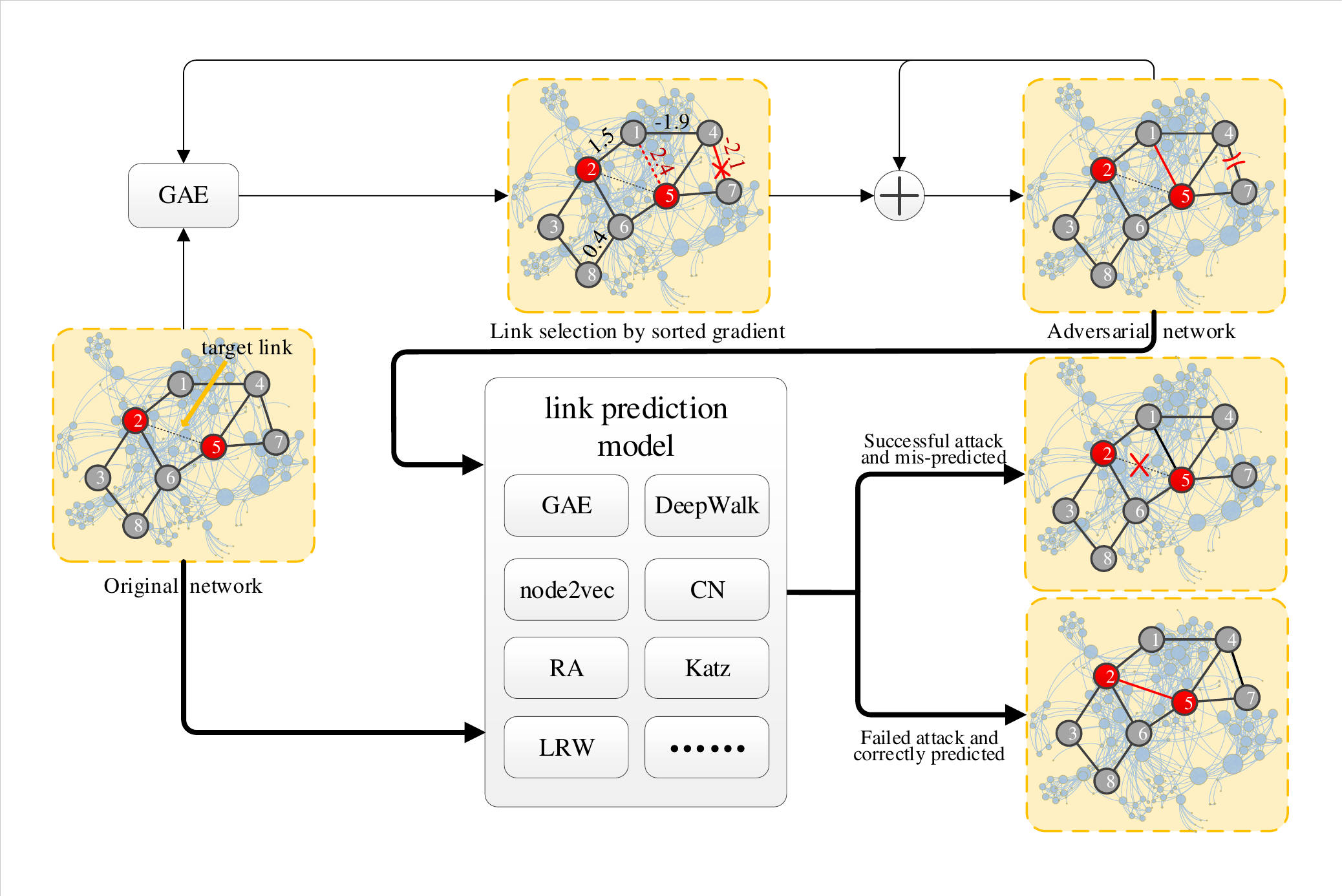}
\caption{The framework of IGA against link prediction methods. First, we choose one link as the target to make predictors cannot precisely work. Then, we calculate gradient for this target to generate a corresponding adversarial network iteratively. Finally, the adversarial network and the original clean network are adopted as the input for various predictors including GAE, DeepWalk, node2vec, CN, RA, Katz and Local Random Walk. These predictors will wrongly predict the target in adversarial network while correctly predict it in the original clean network to implement the adversarial attack}.
 \label{framework}
\end{figure*}

Link prediction adversarial attack is triggered by carefully crafted adversarial network, which is generated via GAE. In this work, the adversarial networks are designed to fool the link prediction methods.
Given observable links of the network as the input, link prediction methods can predict the rest unknown links with satisfying performance.
Then, we generate adversarial networks for hiding the target link to make them mis-predicted. In other words, when the adversarial networks are input, most links can be correctly predicted, while the target link will be mis-predicted. Surprisingly, the adversarial network is almost the same as the original one, and only a few links modified out of thousand links in total compared with the original clean network. We conduct attack through three stages: adversarial network generator, adversarial attack and transferable attack.

In practice, we will probably encounter difficulties when directly implement adversarial attack. Thus, we further design a more feasible attack strategy called single node attack, which can better hide the actions of the attacker. Unlimited attack is also valuable in many scenes where the attacker have high authority. The detail of these two strategies will be discussed below.



\begin{itemize}
\item \textbf{Adversarial network generator.}
 Aiming at the target link and the original network specified, the adversarial network generator is capable of providing the optimal perturbation, effective but unnoticeable. For the target link, we make it as the target of loss function in GAE, and get the gradient network by calculating its partial derivative to the input of GAE. Guided by the gradient information, we get the adversarial network step by step.

\item \textbf{Link prediction adversarial attack.}
Adversarial attack is implemented by adversarial network to lead link prediction method fail to precisely predict the target link. Since adversarial network is generated based on GAE, wrong prediction result of GAE is considered as a successful attack.
In real-world, perturbations added on the clean network may encounter limitations, so we put forward two practical attacks: unlimited attack and single node attack.


\item \textbf{Transferable Attack.}
Lots of excellent link prediction methods are applied to complex networks, and GAE is one of them. The adversarial network generated via GAE is effective against GAE itself, while is it possible that the adversarial attack triggered by the adversarial network can also make other link prediction methods fail as well? We analyze the possibility of transferable adversarial attack and conduct comprehensive experiments to testify the transferability. More interestingly, the attack performance shows that GAE is the most robust method against the adversarial attack compared with DeepWalk~\cite{Perozzi2014DeepWalk}, node2vec~\cite{Grover2016node2vec}, Common Neighbor, Resource Allocation~\cite{Zhou2009Predicting}, Katz index~\cite{Katz1953A} and Local Random Walk~\cite{Liu2010Link}.


\end{itemize}

\subsection{Gradient-based Adversarial Network Generator}
In this section, we explain how to generate the adversarial network based on the gradient information from GAE model.
	
\textbf{GAE model for link prediction.}
Motivated by the convolutional neural network in the computer vision, graph convolutional network (GCN) is proposed to directly implement convolutions on graph~\cite{DBLP:journals/corr/KipfW16}. It has been proved that this propagation rule is the approximation of localized spectral filters on graph. It is worth noting that we use the two-layer GCN to approximate the graph convolution, so that the GAE model can utilize the information of the nodes that are at most 2 hop from the central node.
The embedding vector matrix $Z\in R^{N\times F}$ of each node extracted by GCN layers
is calculated as
\begin{equation}
Z(A)=\hat{A}\sigma(\hat{A}I_NW_{(0)})W_{(1)},
\label{equ:forward}
\end{equation}
where $A$ is the adjacency matrix of the network as the input, $\hat{A}=\tilde{D}^{-\frac{1}{2}}(A+I_N)\tilde{D}^{-\frac{1}{2}}$ is the normalized adjacency matrix, with $I_N$ is the identity matrix also inputted as one-hot feature of node and $\tilde{D}_{ii} = \sum_{j}(A+I_N)_{ij}$
is a diagonal matrix of node degree.
$W_{(0)}\in R^{N\times H}$ and $W_{(1)}\in R^{H\times F}$ represent the weight matrix of the first layer and the second layer of GCN, respectively. $N$ represents the number of nodes, $H$ and $F$ represent the feature dimension of the first and the second layer of GCN. $\sigma$ is the Relu activation function.

After calculating the embedding vector matrix $Z$ for each node,
then we calculate the probability of all links:
\begin{equation}
\tilde{A} = s(ZZ^{T}),
\label{equ:inner-product}
\end{equation}
where $s$ is the sigmoid function and $\tilde{A}\in R^{N\times N}$ is the score matrix. For the link whose score is larger than threshold, we think the link should exist as predicted. In our case, we set the threshold to 0.5.

To train the model, we construct the loss function for supervised training as the cross-entropy error as

\begin{equation}
L=\sum_{ij}- w A_{ij}\ln(\tilde{A}_{ij}) - (1 - A_{ij})\ln(1 - \tilde{A}_{ij}),
\label{equ:loss1}
\end{equation}
where $w=(N^{2}-\sum_{ij}A_{ij})/\sum_{ij}A_{ij}$ is the weight for weighted cross-entropy. Since in a real-world network, the nonexistent links usually is much more than existing links, in other words, the negative sample is much more than positive one, we choose the weighted cross-entropy here as our loss function to prevent the overfitting on the negative samples.

Like many other machine learning tasks, GAE use gradient descent to optimize the parameters in the model
\begin{equation}
W_{i}^{t+1}=W_{i}^{t}-\eta\frac{\partial L}{\partial W_{i}^{t}},
\label{equ:partial}
\end{equation}
where $\eta$ is the learning rate, $i\in\{0, 1\}$ is the layer of weight. $W_{i}^{0}$ is initialized as described in ~\cite{Glorot2010Understanding}.

After properly tuning the hyper-parameters and specific epochs of gradient descent, we can well optimize the parameters of GAE to implement link prediction.

\textbf{Gradient extraction.}
Gradient information is crucial in almost all machine learning tasks, along which we can well optimize the parameters in the model. In our IGA, the gradient is also the direction to add perturbations, but the calculation of gradient here is quite different.
According to Eq.~(\ref{equ:forward}), Eq.~(\ref{equ:inner-product}) and Eq.~(\ref{equ:loss1}), the adjacency matrix is a group of variables in the loss function like the weight matrices $W_{(0)}$ and $W_{(1)}$,
which means we can extract the gradient of adjacency matrix. What differs from the training process is that the loss function in Eq.~(\ref{equ:loss1}) takes all links in $A$ into consideration while for adversarial attack, we just need to consider a single link. For the target link $E_t$, we need construct the target loss function

\begin{equation}
\tilde{L}= - w Y_{t}\ln(\tilde{A}_{t}) - (1 - Y_{t})\ln(1 - \tilde{A}_{t}),
\label{equ:loss2}
\end{equation}
where $Y_{t}\in\{0, 1\}$ is the ground truth label of target link $E_t$ and $\tilde{A}_{t}$ is the probability of the existence of $E_t$ calculated by GAE.

With the loss function specified, we can calculate the partial derivative of $\tilde{L}$ with respect to the adjacency matrix, so as to get the gradient matrix

\begin{equation}
g_{ij}=\frac{\partial \tilde{L}}{\partial A_{ij}}.
\label{equ:partialD}
\end{equation}

Although GAE is a link prediction model only for undirected network, GAE itself actually cannot pick out whether the network we input currently is undirected or not. Therefore, the gradient matrix is usually not symmetry.
Because we focus on the undirected network, we should always symmetrize the gradient matrix. Here, we only keep the upper triangular matrix after symmetrizing

\begin{equation}
\hat{g}_{ij}=\hat{g}_{ji} = \left\{
\begin{array}{ll}
\frac{g_{ij}+g_{ji}}{2} & i < j \\
0 & otherwise.
\end{array}\right.
\label{equ:gradient1}
\end{equation}

\textbf{Adversarial network generator.}
Now, let us consider the detail about how to generate the adversarial network.

In the process of gradient descent, we take the loss $L$ to a tiny value to acquire good prediction ability. Here, on the contrary, we need to maximize the target loss $\tilde{L}$ to let the model wrongly predict the target link. However, due to the discreteness of adjacency matrix, we cannot directly apply gradient ascent on the adjacency matrix to get the adversarial network. Our IGA considers this limitation and extracts the adversarial network as follow.

The value in the gradient matrix could be positive or negative, the positive/negative gradient means the direction of maximizing the target loss is increasing/decreasing the value in the corresponding position of adjacency matrix. Since the network data is discrete, we are only allowed to add or remove links, i.e. we can only add or subtract one in the adjacency matrix. The selection of these links depends on the magnitude of gradient which indicates how significant the link affects the loss function. The larger the magnitude is, the more significantly the link affects the target loss. One thing should be noticed is that due to the discreteness of network data, no matter how large the magnitude is, we are not able to modify a existing/nonexistent link whose gradient is positive/negative. These links is regarded as not attackable and we simply skip them for next attackable link.

Like the process of gradient descent, the generation of adversarial network is also iterative. In every single iteration, we first extract the gradient matrix and symmetrize it to guarantee the network undirected. Then, we choose $n$ links of the largest gradient and is attackable at the same time. By repeat these step for $K$ times, we will get the ultimate adversarial network which can fool
the link prediction methods.

The intact pseuda-code of our attack model is described in Algorithms~1 and~2.

\begin{algorithm}[h]
\caption{Adversarial network generator}
\LinesNumbered
\KwIn{Original network $G$, number of iterations $K$, number of links modified in every iteration $n$.}
\KwOut{The adversarial network $\hat{G}$.}
Train the GAE model on original network $G$\; 
Initialize the adjacency matrix of the adversarial network by $\hat{A}^0=A$\;
\For{h = 1 to $K$}{
	Calculate gradient matrix $g^{h-1}$ based on the $\hat{A}^{h-1}$\;
    Symmetrize $g^{h-1}$ to get $\hat{g}^{h-1}$\;
    $P$ = \textit{Construct perturbation}($\hat{A}^{h-1}$, $\hat{g}^{h-1}$, $n$)\;
    $\hat{A}^{h+1} = \hat{A}^h + P$\;
}
\Return The adversarial network $\hat{G}$, with the adjacency matrix of adversarial network $\hat{A}^K$.\
\end{algorithm}

\begin{algorithm}[h]
\caption{Construct perturbation}
\LinesNumbered
\KwIn{Adjacency matrix $A$, symmetrized gradient matrix $\hat{g}^{h-1}$, number of links modified $n$.}
\KwOut{The perturbation matrix $P$.}
Initialize the perturbation matrix $P$ as a zero matrix with the same size as $A$\;
\For{h = 1 to $n$}{
     Get the position $(i, j)$ of largest magnitude in $\hat{g}^{h-1}$\;
     \uIf{$\hat{g}^{h-1}_{ij} > 0$ and $A_{ij} = 0$}{
     	$P_{ij} = 1$;\	
     }
     \uElseIf{$\hat{g}^{h-1}_{ij} < 0$ and $A_{ij} = 1$}{
     	$P_{ij} = -1$;\	
     }
     \Else{
     	\textbf{Continue}\;
     }
}
$P$ = $P$ + $P^{T}$, where $P^{T}$ is the transpose of P\;
\Return The perturbation matrix $P$.\
\end{algorithm}

\subsection{Adversarial Attack Against GAE}
Since the adversarial network is generated based on GAE, the attack against the GAE model is a typical white-box attack. Although along the direction of gradient, we can generate tiny perturbations, based on which the adversarial network are almost same as the original one. However, under some special circumstances, such tiny perturbations is also hard to implement due to some special limitations and we need to know the performance of our attack model in those cases.
	
    \textbf{Unlimited attack.} Without any limitation on link modification, all the links decided by gradients are valid and the only limitation is the total number of modified links. The applicable scene of unlimited attack is when attacker owns high authority. For instance, a data publisher wants to hide some key links from being detected and there is no doubt that the data publisher has the authority to alter their data freely. Since the number of modified links is well controlled, the data utilities usually will not be damaged.
	
    \textbf{Single node attack.} A target link $E_t$ in the network is a connection of two nodes $(u,v)$, single node attack is defined to modify and only modify links connect to either node $u$ or $v$. We assume in the real-world network, unlimited attack may difficult to conduct since it requires high priority of accessing all information in the network. And in most successful unlimited attacks, the perturbations are usually the modified links that directly connect to node $u$ or $v$.
However, in single node attack, we can promise that at least one of $u$ and $v$ can be well protected without any link modification.
For example, a fraud is the attacker who wants himself to be recommended to an innocent user. In other word, there is no link between the attacker and the innocent user, we conduct attack to lead the recommender predict the link exists.
Instead of directly changing the links of the innocent user node, changing the surroundings, such as links connected to the attacker node, to fool the recommender is more concealed.


\subsection{Transferable Attack}
GAE has overwhelming generalization ability, we assume the adversarial attack can be still effective for other link prediction methods, i.e., the perturbation generated by GAE is universal and the attack thus has strong transferability.
We first test the transferability of adversarial network on unsupervised embedding models including DeepWalk and node2vec, to validate that such attack is quite universal. Then, we conduct the adversarial attack on classic similarity-based indexes for link prediction, such as Common Neighbor, Resource Allocation, Katz index and Local Random Walk, to see whether these traditional methods have the ability to resist the adversarial networks.

The strong transferability of adversarial attacks via IGA may bring the security concern for link prediction applications, since malicious examples may be easily crafted even when the target link prediction method is unknown in advance to conduct effective black-box attacks.


\begin{table}[t]
\centering
\caption{The basic statistics of the three network data sets.}
\label{data sets}
\begin{tabular}{c|ccc}
\hline
\hline
Data Set & \#Nodes & \#Links & \#Average Link Degree  \\ \hline
NS        & 1461  & 2472    & 10.43       \\
Yeast     & 2375 & 11693    & 59.23       \\
Facebook  & 4039 & 88234    & 189.79       \\\hline \hline
\end{tabular}
\end{table}

\begin{table}[t]
\centering
\caption{The link prediction performance comparison of various predictors.}
\label{PredictionPewrformance}
\resizebox{\linewidth}{!}{
\begin{tabular}{c|cc|cc|cc}
\hline\hline
\multirow{3}{*}{Predictors} & \multicolumn{2}{c|}{\multirow{2}{*}{NS}} & \multicolumn{2}{c|}{\multirow{2}{*}{Yeast}} & \multicolumn{2}{c}{\multirow{2}{*}{Facebook}} \\
                            & \multicolumn{2}{c|}{}                    & \multicolumn{2}{c|}{}                       & \multicolumn{2}{c}{}                          \\\cline{2-7}
                            & Accuracy(\%)            & AUC               & Accuracy(\%)              & AUC                & Accuracy(\%)               & AUC                 \\\hline
GAE                         & 100               & 0.9963             & 97.30               & 0.9692             & 99.86                 & 0.9915              \\
DeepWalk                    & 94.16             & 0.9962            & 85.46               & 0.9220              & 92.70                 & 0.9952              \\
node2vec                    & 92.34             & 0.9961            & 87.83               & 0.9224             & 93.06                 & 0.9953              \\
CN                          & 48.18             & 0.9957            & 15.83               & 0.9212             & 31.37                 & 0.9928              \\
RA                          & 77.74             & 0.9962            & 27.29              & 0.9222             & 44.77                 & 0.9953              \\
Katz(0.01)                  & 48.18             & 0.9991            & 26.52               & 0.9711             & 10.01                 & 0.6162              \\
LRW(3)                      & 38.32             & 0.9996            & 14.11               & 0.9732             & 18.60                 & 0.9940               \\
LRW(5)                      & 43.43             & 0.9996            & 10.18               & 0.9771             & 17.16                 & 0.9933 \\\hline \hline
\end{tabular}}
\begin{tablenotes}
\footnotesize
\item[1]It should be noticed that the accuracies of similarity-based methods are much lower than deep learning methods because they are evaluated on $\Omega - E_o$ while deep learning methods only choose a small part from $\Omega - E_o$ for test.
\end{tablenotes}
\end{table}

\begin{table*}[!t]
\centering
\caption{The attack results obtained by different attack methods against different link prediction methods on real-world} data sets.
\label{attack_rst}
\resizebox{\linewidth}{!}{
\begin{tabular}{c|c|cc|ccc|cc|ccc}
\hline \hline
\multicolumn{1}{c|}{\multirow{3}{*}{Data Set}} & \multicolumn{1}{c|}{\multirow{3}{*}{Predictors}} & \multicolumn{5}{c|}{ASR(\%)}                                                       & \multicolumn{5}{c}{AML}                                                 \\ \cline{3-12}
\multicolumn{1}{c|}{}                          & \multicolumn{1}{c|}{}                            & \multicolumn{2}{c|}{IGA}                & \multicolumn{3}{c|}{BASELINE}           & \multicolumn{2}{c|}{IGA}                & \multicolumn{3}{c}{BASELINE} \\ \cline{3-12}
\multicolumn{1}{c|}{}                          & \multicolumn{1}{c|}{}                            & unlimited & \multicolumn{1}{c|}{single} & RAN & DICE & \multicolumn{1}{c|}{GA} & unlimited & \multicolumn{1}{c|}{single} & RAN     & DICE     & GA    \\ \hline
\multirow{9}{*}{NS}   & GAE       &\textbf{56.20}     & 23.72   & 0.00  & 1.82  & 25.18  & \textbf{8.04}  & 17.42 & 10.67  & 11.29 & 9.67  \\
                      & DeepWalk  & 76.83     & \textbf{100}     & 6.67  & 49.81 & 28.52  & 5.74  & \textbf{2.71}  & 11.70  & 9.63  & 9.15  \\
                      & node2vec  & 71.43     & \textbf{96.30}   & 3.33  & 44.44 & 46.67  & 3.36  & \textbf{2.67}  & 12.63  & 4.59  & 3.63  \\
                      & CN        & \textbf{100}       & \textbf{100}     & 2.33  & 83.33 & \textbf{100}    & 4.19  & \textbf{4.18} & 15.02  & 8.17  & 5.16  \\
                      & RA        & 95.71     & 95.31   & 1.37  & 92.02 & \textbf{96.71}  & \textbf{2.80}  & 2.83  & 10.93  & 4.47  & 3.37  \\
                      & Katz(0.01) & \textbf{100}       & \textbf{100}     & 2.27  & 83.33 & \textbf{100}    & \textbf{4.18}  & \textbf{4.18}  & 14.82  & 8.16  & 5.12  \\
                      & LRW(3)    & \textbf{100}       & 98.10   & 2.56  & 92.38 & \textbf{100}    & 2.75  & 2.68  & 8.95   & 3.25  & \textbf{2.56}   \\
                      & LRW(5)    & \textbf{100}     & 99.16   & 4.17   & 94.96   & \textbf{100} &  2.14    & 2.01    & 6.10     & 2.73    & \textbf{2.00}   \\ \cline{2-12}
                      & Average   & 87.52     & \textbf{89.07}   & 2.84  & 67.76 & 74.64  & \textbf{4.15}  & 4.83  & 11.35  & 6.54  & 5.08  \\\hline
\multirow{9}{*}{Yeast}& GAE       & \textbf{69.52} & 32.19 & 0.00 & 2.03  & 36.99 & \textbf{46.78} & 62.65 & 70.87  & 67.27  & 61.21 \\
                      & DeepWalk  & \textbf{96.15} & 92.00 & 3.33 & 76.67 & 86.67 & \textbf{22.77} & 27.20 & 50.27  & 43.60  & 34.53 \\
                      & node2vec  & \textbf{96.15} & 96.00 & 3.33 & 76.67 & 92.31 & \textbf{22.46} & 26.40 & 50.80  & 45.20  & 36.46 \\
                      & CN        & \textbf{100}   & \textbf{100}   & 0.00 & 97.73 & \textbf{100}   & 26.00 & \textbf{21.55} & 141.78 & 33.75  & 35.45\\
                      & RA        & 98.63 & \textbf{100}   & 0.00 & 97.73 & \textbf{100}   & 22.75 & \textbf{19.23} & 70.09  & 42.01  & 32.30\\
                      & Katz(0.01) & \textbf{100}   & \textbf{100}   & 5.00 & 97.85 & \textbf{100}   & \textbf{16.45} & 17.99 & 132.30 & 31.27  & 18.54 \\
                      & LRW(3)    & \textbf{100}   & \textbf{100}   & 0.00 & 96.67 & \textbf{100}   & \textbf{4.10}  & 4.53  & 26.00  & 6.57   & 4.40  \\
                      & LRW(5)    & \textbf{100}   & 95.65 & 0.00 & 95.65 & \textbf{100 }  & 2.78  & \textbf{2.74}  & 28.27  & 3.83   & 2.96  \\\cline{2-12}
                      & Average   & \textbf{95.06} & 89.48 & 1.46 & 80.13 & 89.50 & \textbf{20.51} & 22.79 & 71.30  & 34.19  & 28.23 \\ \hline
\multirow{9}{*}{Facebook}& GAE    & \textbf{52.84} & 22.74& 0.00 & 0.33  & 15.05 &\textbf{ 134.97}& 171 .77& 198.57 & 189.94 & 161.51 \\
                      &DeepWalk   & \textbf{100}   & \textbf{100}  & 0.00 & \textbf{100}   & \textbf{100}   & 99.26 & \textbf{79.00}  & 160.71 & 93.33  & 135.00 \\
                      &node2vec   & \textbf{100}   & \textbf{100}  & 0.00 & \textbf{100}   & \textbf{100}   & \textbf{80.10} & 83.67  & 176.00 & 94.33  & 132.00 \\
                      & CN        & \textbf{100}   & \textbf{100}  & 3.12 & 98.92 & \textbf{100}   & 49.78 & \textbf{45.72}  & 290.94 & 70.09  & 54.23  \\
                      & RA        & \textbf{100}   & \textbf{100}  & 0.00 & 90.77 & 94.57 & 40.50 & \textbf{35.38}  & 264.56 & 81.43  & 72.85  \\
                      &Katz(0.01)  & \textbf{100}   & \textbf{100}  & 0.00 & 81.25 & \textbf{100}   & 55.63 & \textbf{53.67}  & 204.00 & 114.44 & 61.06  \\
                      & LRW(3)    & \textbf{100}   & \textbf{100}  & 11.11& 93.22 & 91.38 & \textbf{15.98} & 22.31  & 232.11 & 61.12  & 25.50  \\
                      & LRW(5)    & \textbf{100}   & \textbf{100}  & 11.11& 94.44 & 96.23 & \textbf{16.19} & 19.67  & 232.11 & 66.98  & 20.85  \\\cline{2-12}
                      & Average   & \textbf{94.11} & 90.34& 3.17 & 82.37 & 87.15 & \textbf{61.55} & 63.90  & 219.87 & 96.46    & 82.87   \\\hline \hline
\end{tabular}}

\begin{tablenotes}
\footnotesize
\item[1]For the Katz index, we set the damping factor to 0.01.
\item[2]For two Local Random Walk predictors, we set the total walk steps to 3 and 5 respectively.
\end{tablenotes}
\end{table*}

\section{Experiments\label{Exp}}
In this section, we conduct adversarial attack on the GAE compared with three baseline methods. Then we use the adversarial network to test the robustness of a series of link prediction methods. Further, we check the performance of IGA under different limitations. 

\subsection{Experiments Setup}
The data sets, baselines, parameter setting and link prediction algorithms for transferability testing will be introduced here. Our experimental environment consists of i7-7700K 3.5GHzx8 (CPU), TITAN Xp 12GiB (GPU), 16GBx4 memory (DDR4) and Ubuntu 16.04 (OS).

\textbf{Data sets.}
To test the performance of our IGA, We choose three networks of different types for adversarial attack. The data sets are all undirected and unweighted with different scale.

\begin{itemize}
\item \textbf{NS}~\cite{Newman2006Finding}: a network of co-authorships between scientists who published papers in the field of network. It contains 1589 scientists, 128 of which are isolated. We simply remove those isolated nodes since prediction on those nodes are meaningless.

\item \textbf{Yeast}~\cite{Von2002Comparative}: a protein-protein interaction network whose giant component containing 2375 proteins and 11693 interactions. Although the network is not well-connected, the giant component contains the 90.75\% of all nodes and we choose it for our experiment.

\item \textbf{Facebook}~\cite{Mcauley2012Learning}: a social network where nodes represent the users and links are their friendship relations. Facebook is a large scale network containing 4039 nodes and 88234 links.
\end{itemize}
	
We summarize the topological feature of these data sets shown in the TABLE~\ref{data sets}.

Unlimited attack and single node attack are conducted on these data sets.

\textbf{Baseline.}
We compare IGA with three baseline methods. Since this is the first time link prediction adversarial attack is proposed, there's no attack methods for adversarial attack against link prediction methods. For comparison, we modified attack methods to implement link prediction attack.

\begin{itemize}
\item \textbf{Random Attack (RAN)}: RAN randomly disconnects $a$ links in the original network, while randomly connects $b$ pairs of nodes that are originally not connected. This is the simplest attack method.

\item \textbf{Disconnect Internally, Connect Externally (DICE)}~\cite{waniek2018hiding}:
DICE originally is a heuristic algorithm for disguising communities. For the target community in the network, DICE firstly delete $a$ inner links and connected $b$ links from the community to nodes in a distance to make the target invisible. In our experiment, we replace the community by the target link, and the internal links are those links directly connecting to the target link.


\item \textbf{Gradient Attack (GA)}:
GA is a gradient based adversarial attack method without iterations, i.e. GA generates perturbations by just calculating gradient once. Therefore, the computational complexity of GA is much lower than IGA while the accuracy is much lower as a tradeoff. To realize GA, we just need to set the parameter $K$=$1$ in IGA.
\end{itemize}



\textbf{Parameter Setting.}
The first parameter in our experiment is the number of modified link in the adversarial network. Intuitively, if we set it very large, the goal of fooling the predictor will easily achieve but the neighbors of the target link will totally change. Because the perturbations are too noticeable, the attack itself consequently is pointless. How to set the number of modified link actually indicates how to define the unnoticeable perturbations.
Because modified links usually directly connect to the target link, in this paper, we think the perturbation size should relate to the degree of target link $k_t$ and should be different for different target.
The link degree, a.k.a. edge degree, is defined as the number of links connected to this link.
For nonexistent link, the link degree is the summation of the degree of two nodes that form this link. For existing link, we should exclude the link itself when sum the degree.
If the number of modified links is much smaller than the target link’s degree, we think the perturbations can be regarded as unnoticeable.
Therefore, in our experiments, the number of changed links of three methods is set to the target link degree $k_t$.

For IGA, we set  $n=1$ for highest accuracy and consequently, set $K=k_t$ because $n\times K$ actually is the number of modified link in total.
In the baseline DICE and RAN, we set $a=k_t/2$ and $b=k_t/2$.

\textbf{Evaluation metric.}
To evaluate the performance of attack, we use two metrics introduced as follow.

\begin{itemize}

\item \textbf{Attack Success Rate (ASR)}:
The attack success rate, i.e., the ratio of the target links which will be successfully attack within $k_t$ modified links for each target versus all target links. The larger ASR corresponds to the better attack effect.
\item \textbf{Average Modified Link number (AML)}:
It denotes the average perturbation size leading to successful attack. Although the total perturbation size for a target is set to target link degree $k_t$, we still focus on the minimal perturbation size making attack successful.
For those failed attack, we regard it as the total perturbation size $k_t$. The smaller the AML is, the better the attack performance is.
\end{itemize}




\begin{figure*}[!t]
  \centering
\includegraphics[width=1\linewidth]{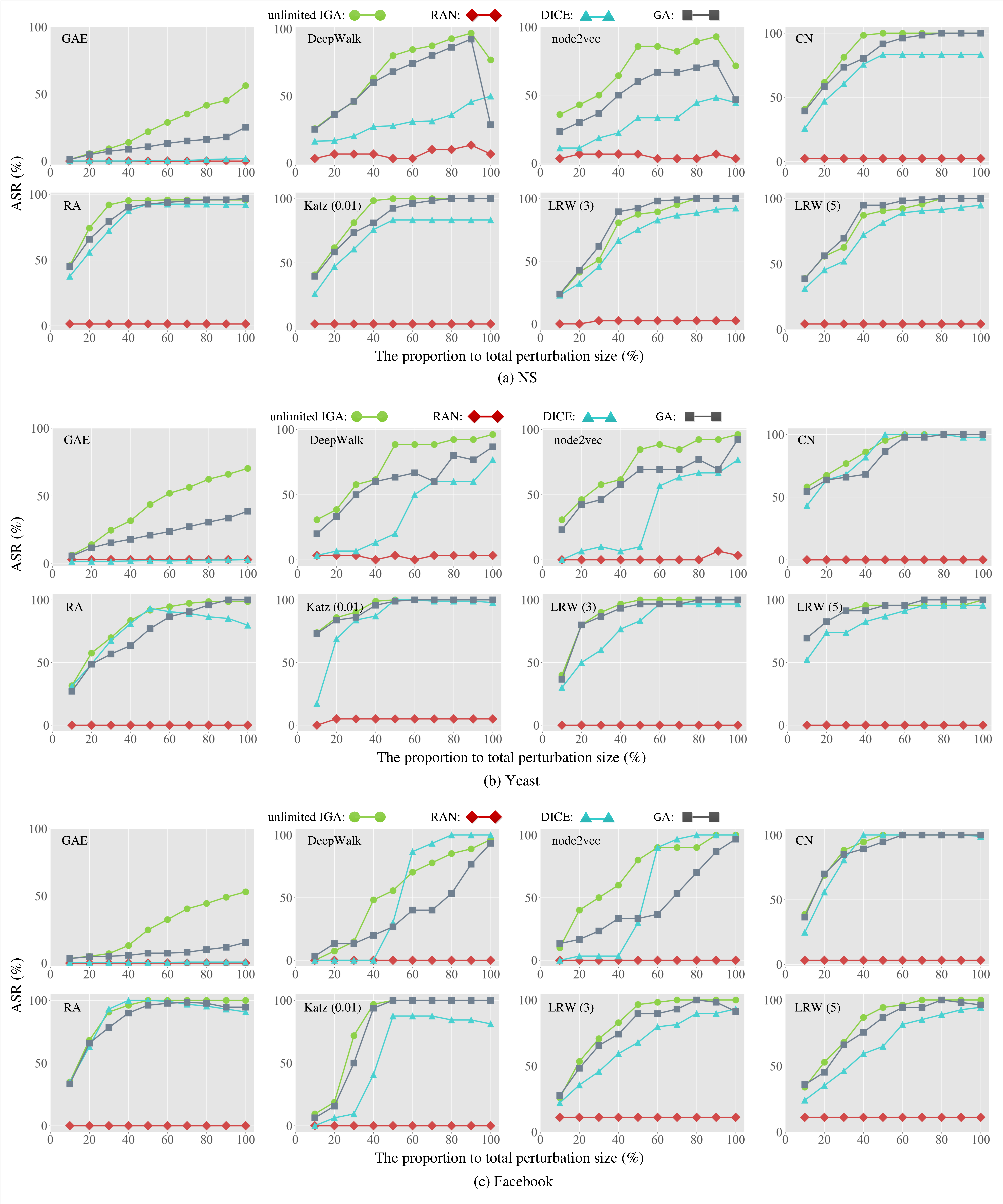}
\caption{Different attack methods against various of link prediction methods}
 \label{fig}
\end{figure*}

\subsection{Results}
Comprehensive experiments are carried out to testify the attack performance and transfer attack. In this part, we will show our experimental result in different aspects how IGA is efficient compared with other baseline methods against a bunch of link prediction methods on different real-world data sets.

\textbf{IGA has the best overall performance.}
We present the overall attack success rate and perturbations of adversarial network in TABLE~\ref{attack_rst}. It is obvious that IGA has the best overall performance. In most cases, compared with baseline attack methods, IGA is the best attack method with highest attack success rate and least average modified links in adversarial networks. Surprisingly, even though the perturbation size is not above the link degree, the ASR of IGA against most predictors is exactly, or very close to 100\%, except the attack on GAE itself. The GA, whose computational complexity is much less than IGA, still outperforms DICE with the ASR being 100\% against most similarity-based predictors. As to the attack overhead, the AML of IGA is significantly less than other attack baselines. Especially, for the Facebook network, the average AML of IGA is 40\% less than DICE, which indicates that the adversarial network generated by IGA is much more concealed and the gradient is an accurate direction to generate perturbation.

\textbf{Robustness analysis of networks.}
Impressively, for the network Yeast and Facebook, although the average link degree is quite large, we can still successfully conduct adversarial attack with very few links changed, i.e. the ratio of average AML to average link degree is smaller than 35\%. Intuitively, for the network with dense structure, it will be more robust against adversarial attack. However, from our experimental results, we can see that the average ASR on Yeast and Facebook is higher than NS. Since the link in dense network has more complex environments, the modified links could be more unnoticeable, which is concerning.

\textbf{GAE is the most robust predictor against adversarial attack.}
Generally, white-box attack has better attack capacity than black-box attack, since white-box has grasped the internal information of the target methods, which is easier to trigger effective attack. Interestingly, in our case, the experiments show that black-box attack against traditional link prediction methods has much higher success rate than against GAE itself.
Take NS and Facebook for examples, the attack success rate against GAE is no higher than 60\%, while attack against similarity-based indexes is as high as 100\%. Especially for the Local Random Walk, IGA can successfully fool it by only 3 or 4 links modified out of ten thousand in Yeast, whose average link degree is about 60. The poor resistance of those classic link prediction methods to adversarial attack indicates that the robustness of them should be highly concerned. Usually, the vulnerability of deep model is concerned the most, but according to the comprehensive experiments results, we find that GAE is the most robust one. The DeepWalk and node2vec are also deep models, and they can better resist the adversarial attack than similarity-based methods.

\textbf{Single node attack outperforms unlimited attack.}
We compare the ASR of single node attack and unlimited attack in TABLE~\ref{attack_rst}. Another interesting note is that although single node attack is a constrained conditional attack which limit its modification target in network, however, it still outperforms unlimited attack. For example, in the NS data set, the ASR of DeepWalk and node2vec under single node attack is close to 100\% while the ASR under unlimited attack is less than 80\%. We can explain the results. In single node attack, the modified link is more focused on the one target node, and the neighbors of the node may change more significantly than unlimited attack. Therefore, although another node is well protected, the goal of fooling predictors still can easily achieve.

\textbf{The visualization of four attack methods.}
In Fig.~\ref{fig}, we visualize the four attack methods against eight link prediction methods on three data sets. The attack against GAE shows that IGA is much better than GA in white-box attack because of the high accuracy of iteratively calculating gradient. IGA also outperforms DICE especially when against deep learning methods GAE, DeepWalk and node2vec in NS and Yeast. But as to the Facebook data set, DICE performs better than our IGA when against DeepWalk and node2vec. When the perturbation size is above 50\%, DICE starts to add links which connect the target to nodes in a distance. We think since the Facebook network is quite large, the embedding generated by random walk could change quite significantly if some nodes in a distance was connected to the target. For the similarity-based methods, the performance of all attack methods except RAND is close because of the vulnerability of these link prediction methods. In most cases, obviously, we still can find that IGA and GA are better than DICE especially when the perturbation is under 50\%.

\textbf{The attack on links of different degree.}
TABLE.~\ref{degree} shows the statistical result of ASR on links with different degree. Since the Facebook network is much denser than NS and Yeast, the ASR where only 3 links modified are all zero. Therefore, we only post the experimental results of NS and Yeast. Surprisingly, with less than 3 links modified, a large number of targets will be successfully attack especially for those with low degree. For example, with only 3 links modified, the ASR are 33.93\% and 19.15\% for link in Yeast whose degree are in range 3 to 20 and 21 to 30 respectively.

\begin{table}[]
\caption{ASR of links with different degree.}
\label{degree}
\centering
\resizebox{\linewidth}{!}{
\begin{tabular}{c|c|ccccccc}
\hline \hline
\multirow{2}{*}{Networks} & \multirow{2}{*}{Modified links} & \multicolumn{7}{c}{Degree}                                                                                                                                                                                                         \\ \cline{3-9}
                          &                                 & \multicolumn{1}{l}{{[}2-3{]}} & \multicolumn{1}{l}{{[}4-5{]}} & \multicolumn{1}{l}{{[}6-7{]}} & \multicolumn{1}{l}{{[}8-12{]}} & \multicolumn{1}{l}{{[}13-20{]}} & \multicolumn{1}{l}{{[}21-30{]}} & \multicolumn{1}{l}{{[}31-{]}} \\ \hline
\multirow{3}{*}{NS}       & 1  & 0.00    & 0.00    & 0.00   & 0.00    & 0.00   & 0.00   & 0.00 \\
                          & 2  & 35.14   & 5.00    & 1.96   & 5.88    & 8.77   & 7.14   & 8.33  \\
                          & 3  & 67.57   & 22.5    & 17.67  & 19.51   & 14.04  & 11.76  & 9.52   \\ \hline
\multirow{3}{*}{Yeast}    & 1  & 12.5    & 4.26    & 1.43   & 2.78    & 0.00   & 0.00   & 0.00  \\
                          & 2  & 23.21   & 10.64   & 7.14   & 2.78    & 0.00   & 0.00   & 0.00   \\
                          & 3  & 33.93   & 19.15   & 10.00  & 2.78    & 0.00   & 0.00   & 0.00  \\ \hline \hline

\end{tabular}}
\end{table}


\section{Conclusion\label{Conclusion}}
In this work, we propose a novel adversarial attack model IGA against link prediction methods. Our IGA is based on GAE link prediction methods, but experimental results shows that IGA is quite effective to conduct adversarial on a various of link prediction methods including deep learning methods and classic similarity-based methods. IGA can be utilized as the method for privacy protection or the metric for robustness evaluation of link prediction methods. Since we should take the whole adjacency matrix as the input for gradient calculation, the adversarial attack on larger scale network is quite difficult due to the memory limitation. Further work should focus on solving this problem. The further study of how to defend such link prediction adversarial attack is also of great importance.

\ifCLASSOPTIONcompsoc
  \section*{Acknowledgments}
\else
  \section*{Acknowledgment}
\fi

This work is partially supported by National Natural Science Foundation of China (61502423, 61572439), Zhejiang Science and Technology Plan Project (LGF18F030009), Zhejiang University of Technology Control Science and Engineering Open Fund.


\ifCLASSOPTIONcaptionsoff
  \newpage
\fi




\bibliographystyle{IEEEtran}
\bibliography{ref1802}

\begin{thebibliography}{10}
\providecommand{\url}[1]{#1}
\csname url@samestyle\endcsname
\providecommand{\newblock}{\relax}
\providecommand{\bibinfo}[2]{#2}
\providecommand{\BIBentrySTDinterwordspacing}{\spaceskip=0pt\relax}
\providecommand{\BIBentryALTinterwordstretchfactor}{4}
\providecommand{\BIBentryALTinterwordspacing}{\spaceskip=\fontdimen2\font plus
\BIBentryALTinterwordstretchfactor\fontdimen3\font minus
  \fontdimen4\font\relax}
\providecommand{\BIBforeignlanguage}[2]{{%
\expandafter\ifx\csname l@#1\endcsname\relax
\typeout{** WARNING: IEEEtran.bst: No hyphenation pattern has been}%
\typeout{** loaded for the language `#1'. Using the pattern for}%
\typeout{** the default language instead.}%
\else
\language=\csname l@#1\endcsname
\fi
#2}}
\providecommand{\BIBdecl}{\relax}
\BIBdecl

\bibitem{Borgatti2009Network}
S.~P. Borgatti, A.~Mehra, D.~J. Brass, and G.~Labianca, ``Network analysis in
  the social sciences,'' \emph{Science}, vol. 323, no. 5916, pp. 892--895,
  2009.

\bibitem{Wellman2005The}
B.~Wellman, ``The development of social network analysis: A study in the
  sociology of science by linton c. freeman,'' \emph{Social Networks}, vol.~27,
  no.~3, pp. 275--282, 2005.

\bibitem{Montoya2002Small}
J.~M. Montoya and R.~V. Sol, ``Small world patterns in food webs.''
  \emph{Journal of Theoretical Biology}, vol. 214, no.~3, pp. 405--412, 2002.

\bibitem{Ebel2002Scale}
H.~Ebel, L.~I. Mielsch, and S.~Bornholdt, ``Scale-free topology of e-mail
  networks,'' \emph{Phys Rev E Stat Nonlin Soft Matter Phys}, vol.~66, no. 3 Pt
  2A, p. 035103, 2002.

\bibitem{Latora2002Is}
V.~Latora and M.~Marchiori, ``Is the boston subway a small-world network?''
  \emph{Physica A Statistical Mechanics \& Its Applications}, vol. 314, no.~1,
  pp. 109--113, 2002.

\bibitem{Neal2008}
J.~W. Neal, ``"kracking" the missing data problem: Applying krackhardt's
  cognitive social structures to school-based social networks,''
  \emph{Sociology of Education}, vol.~81, no.~2, pp. 140--162, 2008.

\bibitem{Liben2003The}
D.~Liben‐Nowell and J.~Kleinberg, ``The link‐prediction problem for social
  networks,'' 2003, pp. 556--559.

\bibitem{Anil2016Link}
A.~Anil, D.~Kumar, S.~Sharma, R.~Singha, R.~Sarmah, N.~Bhattacharya, and S.~R.
  Singh, ``Link prediction using social network analysis over heterogeneous
  terrorist network,'' in \emph{IEEE International Conference on Smart
  City/socialcom/sustaincom}, 2016, pp. 267--272.

\bibitem{chen2017improved}
J.~Chen, Y.~Wu, L.~Fan, X.~Lin, H.~Zheng, S.~Yu, and Q.~Xuan, ``Improved
  spectral clustering collaborative filtering with node2vec technology,'' in
  \emph{Complex Systems and Networks (IWCSN), 2017 International Workshop
  on}.\hskip 1em plus 0.5em minus 0.4em\relax IEEE, 2017, pp. 330--334.

\bibitem{chen2017double}
J.~Chen, X.~Lin, Y.~Wu, Y.~Chen, H.~Zheng, M.~Su, S.~Yu, and Z.~Ruan, ``Double
  layered recommendation algorithm based on fast density clustering: Case study
  on yelp social networks dataset,'' in \emph{Complex Systems and Networks
  (IWCSN), 2017 International Workshop on}.\hskip 1em plus 0.5em minus
  0.4em\relax IEEE, 2017, pp. 242--252.

\bibitem{Huang2005Link}
Z.~Huang, ``Link prediction approach to collaborative filtering,'' pp.
  141--142, 2005.

\bibitem{Yang2012Scalable}
X.~Yang, Z.~Zhang, and K.~Wang, ``Scalable collaborative filtering using
  incremental update and local link prediction,'' in \emph{Proceedings of the
  21st ACM international conference on Information and knowledge
  management}.\hskip 1em plus 0.5em minus 0.4em\relax ACM, 2012, pp.
  2371--2374.

\bibitem{Guimer2009Missing}
R.~Guimerà and M.~Salespardo, ``Missing and spurious interactions and the
  reconstruction of complex networks,'' \emph{Proceedings of the National
  Academy of Sciences of the United States of America}, vol. 106, no.~52, pp.
  22\,073--22\,078, 2009.

\bibitem{Bilgic2007Combining}
M.~Bilgic, G.~M. Namata, and L.~Getoor, ``Combining collective classification
  and link prediction,'' in \emph{IEEE International Conference on Data Mining
  Workshops, 2007. ICDM Workshops}, 2007, pp. 381--386.

\bibitem{Lin1998An}
D.~Lin, ``An information-theoretic definition of similarity,'' in
  \emph{International Conference on Machine Learning}, 1998, pp. 296--304.

\bibitem{Popescul2003Statistical}
A.~Popescul, ``Statistical relational learning for link prediction,'' in
  \emph{Proc. IJCAI Workshop on Learning Statistical MODELS From Relational
  Data}, 2003.

\bibitem{Craven1998Learning}
M.~Craven, D.~Dipasquo, D.~Freitag, A.~Mccallum, T.~Mitchell, K.~Nigam, and
  S.~Slattery, ``Learning to extract symbolic knowledge from the world wide
  web,'' in \emph{Proc. of the National Conference on Artificial Intelligence},
  1998, pp. 121--126.

\bibitem{Zhou2009Predicting}
T.~Zhou, L.~Lü, and Y.~C. Zhang, ``Predicting missing links via local
  information,'' \emph{European Physical Journal B}, vol.~71, no.~4, pp.
  623--630, 2009.

\bibitem{Zhou2007Bipartite}
T.~Zhou, J.~Ren, M.~Medo, and Y.~C. Zhang, ``Bipartite network projection and
  personal recommendation.'' \emph{Physical Review E Statistical Nonlinear \&
  Soft Matter Physics}, vol.~76, no.~2, p. 046115, 2007.

\bibitem{L2011Link}
L.~Lü and T.~Zhou, ``Link prediction in complex networks: A survey,''
  \emph{Physica A Statistical Mechanics \& Its Applications}, vol. 390, no.~6,
  pp. 1150--1170, 2011.

\bibitem{Liu2010Link}
W.~Liu and L.~Lu, ``Link prediction based on local random walk,'' vol.~89,
  no.~5, pp. 58\,007--58\,012(6), 2010.

\bibitem{Katz1953A}
L.~Katz, ``A new status index derived from sociometric analysis,''
  \emph{Psychometrika}, vol.~18, no.~1, pp. 39--43, 1953.

\bibitem{Mikolov2013Efficient}
T.~Mikolov, K.~Chen, G.~Corrado, and J.~Dean, ``Efficient estimation of word
  representations in vector space,'' \emph{Computer Science}, 2013.

\bibitem{Perozzi2014DeepWalk}
B.~Perozzi, R.~Al-Rfou, and S.~Skiena, ``Deepwalk:online learning of social
  representations,'' in \emph{ACM SIGKDD International Conference on Knowledge
  Discovery and Data Mining}, 2014, pp. 701--710.

\bibitem{Tang2015LINE}
J.~Tang, M.~Qu, M.~Wang, M.~Zhang, J.~Yan, and Q.~Mei, ``Line:large-scale
  information network embedding,'' vol.~2, no.~2, pp. 1067--1077, 2015.

\bibitem{Grover2016node2vec}
A.~Grover and J.~Leskovec, ``node2vec: Scalable feature learning for
  networks,'' in \emph{ACM SIGKDD International Conference on Knowledge
  Discovery and Data Mining}, 2016, pp. 855--864.

\bibitem{Bunke2008Graph}
H.~Bunke and K.~Riesen, ``Graph classification based on dissimilarity space
  embedding,'' in \emph{Joint Iapr International Workshop on Structural,
  Syntactic, and Statistical Pattern Recognition}, 2008, pp. 996--1007.

\bibitem{Riesen2010Graph}
K.~Riesen and H.~Bunke, ``Graph classification and clustering based on vector
  space embedding,'' vol. Volume 77, p. 348, 2010.

\bibitem{Tang2015PTE}
J.~Tang, M.~Qu, and Q.~Mei, ``Pte: Predictive text embedding through
  large-scale heterogeneous text networks,'' in \emph{ACM SIGKDD International
  Conference on Knowledge Discovery and Data Mining}, 2015, pp. 1165--1174.

\bibitem{Wang2016Linked}
S.~Wang, J.~Tang, C.~Aggarwal, and H.~Liu, ``Linked document embedding for
  classification,'' in \emph{CIKM}, 2016.

\bibitem{Tian2014Learning}
F.~Tian, B.~Gao, Q.~Cui, E.~Chen, and T.~Y. Liu, ``Learning deep
  representations for graph clustering,'' in \emph{Twenty-Eighth AAAI
  Conference on Artificial Intelligence}, 2014, pp. 1293--1299.

\bibitem{Allab2016A}
K.~Allab, L.~Labiod, and M.~Nadif, ``A semi-nmf-pca unified framework for data
  clustering,'' \emph{IEEE Transactions on Knowledge \& Data Engineering},
  vol.~29, no.~1, pp. 2--16, 2016.

\bibitem{Kipf2016Variational}
T.~N. Kipf and M.~Welling, ``Variational graph auto-encoders,'' 2016.

\bibitem{Schlichtkrull2017Modeling}
M.~Schlichtkrull, T.~N. Kipf, P.~Bloem, R.~V.~D. Berg, I.~Titov, and
  M.~Welling, ``Modeling relational data with graph convolutional networks,''
  2017.

\bibitem{Goodfellow2014Explaining}
I.~J. Goodfellow, J.~Shlens, and C.~Szegedy, ``Explaining and harnessing
  adversarial examples,'' \emph{Computer Science}, 2014.

\bibitem{Szegedy2013Intriguing}
C.~Szegedy, W.~Zaremba, I.~Sutskever, J.~Bruna, D.~Erhan, I.~Goodfellow, and
  R.~Fergus, ``Intriguing properties of neural networks,'' \emph{Computer
  Science}, 2013.

\bibitem{Moosavidezfooli2016DeepFool}
S.~M. Moosavidezfooli, A.~Fawzi, and P.~Frossard, ``Deepfool: A simple and
  accurate method to fool deep neural networks,'' in \emph{Computer Vision and
  Pattern Recognition}, 2016, pp. 2574--2582.

\bibitem{nagaraja2010impact}
S.~Nagaraja, ``The impact of unlinkability on adversarial community detection:
  effects and countermeasures,'' in \emph{International Symposium on Privacy
  Enhancing Technologies Symposium}.\hskip 1em plus 0.5em minus 0.4em\relax
  Springer, 2010, pp. 253--272.

\bibitem{waniek2018hiding}
M.~Waniek, T.~P. Michalak, M.~J. Wooldridge, and T.~Rahwan, ``Hiding
  individuals and communities in a social network,'' \emph{Nature Human
  Behaviour}, vol.~2, no.~2, p. 139, 2018.

\bibitem{DBLP:conf/kdd/ZugnerAG18}
D.~Z{\"{u}}gner, A.~Akbarnejad, and S.~G{\"{u}}nnemann, ``Adversarial attacks
  on neural networks for graph data,'' in \emph{Proceedings of the 24th {ACM}
  {SIGKDD} International Conference on Knowledge Discovery {\&} Data Mining,
  {KDD} 2018, London, UK, August 19-23, 2018}, 2018, pp. 2847--2856.

\bibitem{bojcheski2018adversarial}
A.~Bojchevski and S.~G{\"u}nnemann, ``Adversarial attacks on node embeddings,''
  \emph{arXiv preprint arXiv:1809.01093}, 2018.

\bibitem{dai2018adversarial}
H.~Dai, H.~Li, T.~Tian, X.~Huang, L.~Wang, J.~Zhu, and L.~Song, ``Adversarial
  attack on graph structured data,'' \emph{arXiv preprint arXiv:1806.02371},
  2018.

\bibitem{chen2018fast}
J.~Chen, Y.~Wu, X.~Xu, Y.~Chen, H.~Zheng, and Q.~Xuan, ``Fast gradient attack
  on network embedding,'' \emph{arXiv preprint arXiv:1809.02797}, 2018.

\bibitem{zheleva2008preserving}
E.~Zheleva and L.~Getoor, ``Preserving the privacy of sensitive relationships
  in graph data,'' in \emph{Privacy, security, and trust in KDD}.\hskip 1em
  plus 0.5em minus 0.4em\relax Springer, 2008, pp. 153--171.

\bibitem{fard2012limiting}
A.~M. Fard, K.~Wang, and P.~S. Yu, ``Limiting link disclosure in social network
  analysis through subgraph-wise perturbation,'' in \emph{Proceedings of the
  15th International Conference on Extending Database Technology}.\hskip 1em
  plus 0.5em minus 0.4em\relax ACM, 2012, pp. 109--119.

\bibitem{fard2015neighborhood}
A.~M. Fard and K.~Wang, ``Neighborhood randomization for link privacy in social
  network analysis,'' \emph{World Wide Web}, vol.~18, no.~1, pp. 9--32, 2015.

\bibitem{DBLP:journals/corr/KipfW16}
\BIBentryALTinterwordspacing
T.~N. Kipf and M.~Welling, ``Semi-supervised classification with graph
  convolutional networks,'' \emph{CoRR}, vol. abs/1609.02907, 2016. [Online].
  Available: \url{http://arxiv.org/abs/1609.02907}
\BIBentrySTDinterwordspacing

\bibitem{Glorot2010Understanding}
X.~Glorot and Y.~Bengio, ``Understanding the difficulty of training deep
  feedforward neural networks,'' \emph{Journal of Machine Learning Research},
  vol.~9, pp. 249--256, 2010.

\bibitem{Newman2006Finding}
M.~E. Newman, ``Finding community structure in networks using the eigenvectors
  of matrices,'' \emph{Physical Review E Statistical Nonlinear \& Soft Matter
  Physics}, vol.~74, no. 3 Pt 2, p. 036104, 2006.

\bibitem{Von2002Comparative}
M.~C. Von, R.~Krause, B.~Snel, M.~Cornell, S.~G. Oliver, S.~Fields, and
  P.~Bork, ``Comparative assessment of large-scale data sets of protein-protein
  interactions.'' \emph{Nature}, vol. 417, no. 6887, pp. 399--403, 2002.

\bibitem{Mcauley2012Learning}
J.~Mcauley and J.~Leskovec, ``Learning to discover social circles in ego
  networks,'' in \emph{International Conference on Neural Information
  Processing Systems}, 2012, pp. 539--547.

\end{thebibliography}

\vfill




\end{document}